\documentstyle[preprint,prl,aps]{revtex}
\tightenlines
\begin{document}
\draft
\preprint{LV6382}
\title{A statistical mechanics of an oscillator associative memory 
with scattered natural frequencies 
} 
\author{Toru Aonishi \and Koji Kurata}
\address{
Department of Systems and Human Science, 
Osaka University, 1-3, Toyonaka, Osaka 560, Japan
}

\author{Masato Okada}
\address{ERATO Kawato Dynamic Brain Project, 2-2 Hikaridai, Seika-cho, 
Soraku-gun, Kyoto 619-02, Japan
}

\date{\today}
\maketitle

\begin{abstract}
Analytic treatment of a non-equilibrium random system
with large degrees of freedoms is
one of most important problems of physics.
However, little research has been done on this problem
as far as we know.
In this paper, we propose a new mean field theory 
that can treat a general class of a non-equilibrium random system.
We apply the present theory to an
analysis for an associative memory with oscillatory elements, 
which is a well-known typical random system
with large degrees of freedoms.
\end{abstract}

\narrowtext

The analytic treatment of a non-equilibrium random system
with large degrees of freedoms is
one of most important problems of physics.
However, little research has been done on this problem
as far as we know.
In this paper, we propose a new mean field theory 
that can treat a general class of a non-equilibrium random system.
We apply the present theory to an
analysis for an associative memory with oscillatory elements, 
which is a well-known typical random system
with large degrees of freedoms.

Historically,
there are two important studies
regarding the non-equilibrium systems with large degrees of
freedoms that cannot be treated
by conventional equilibrium statistical 
mechanics, since we can not define the Lyapunov function 
with ``bottoms''. 
Kuramoto \cite{kuramoto} theoretically analyzed
the mutual entrainment of uniformly coupled oscillators 
with scattered natural frequencies (Kuramoto theory). 
His model corresponds 
to a mean field model of a ferromagnet
in equilibrium statistical mechanics.
Kuramoto utilized the spilite of statistical mechanics,
namely, a notion of macroscopic order-parameters,
to investigate his non-equilibrium system
with large degrees of freedoms.
Daido \cite{daido} numerically analyzed the quasi-entrainment 
of randomly coupled oscillators with scattered natural frequencies.
His model corresponds to 
the Sherrington-Kirkpatrick model of a spin glass \cite{sherrigton}
in equilibrium statistical mechanics.
A mean field theory should be developed 
for non-equilibrium random systems {\it with} frustration
that follows the history of mean field theories
in the equilibrium statistical mechanics.
However, this kind of non-equilibrium random system has not yet
been theoretically analyzed 
(see \cite{daido2} for the random system {\it without} frustration). 
This is a famous ``open 
problem'' in physics \cite{science}.
That is reason why we have proposed the present theory.

On the other hand, the Lyapunov function can not be defined 
in spin-like systems with the non-monotonic reaction function either.
In the self-consistent signal-to-noise analysis (SCSNA) \cite{fukai3}, 
the notion of a macroscopic order parameter
is also introduced to analyze these frustrated spin-like systems,
which cannot be treated by conventional equilibrium statistical mechanics.
Note that the results of applying the SCSNA 
to  simple random spin systems \cite{fukai3,okada}, i.e., 
the Sherrington-Kirkpatrick model and the Hopfield model \cite{hopfield},  
coincide with those of the replica theories
\cite{Bray1986,Kuhn1991}.

In this paper, we propose a new theoretical framework for
an oscillator associative memory model with
scattered natural frequencies in memory retrieval states. 
This system can be considered as a typical example
of non-equilibrium random systems with large degrees of freedoms.
The present theory makes a bridge between 
the SCSNA and the Kuramoto theory.
Using the same procedure, we can easily treat 
a glass oscillator system \cite{daido}.
Our theory is reduced to the Kuramoto theory in the finite loading case.
When all oscillators have uniform natural frequencies, 
our theory coincides with the previously proposed theories 
\cite{okuda,cook} 
in the equilibrium statistical mechanics
for an XY spin system.

The mutual entrainment is an important notion 
of a non-equilibrium system with large degrees of freedoms.
In uniformly coupled oscillators, there is 
a unique stable state, the ferromagnetic phase in the phase space.  
On the other hand, in frustrated systems, there are  
many stable states in the phase space. 
We need to elucidate the properties of the mutual entrainment
in each stable phase (ferromagnetic phase and glass phase).
Our new theory describes a phenomenon of the mutual entrainment 
in the ferromagnetic phase (memory retrieval). 
Thus, we numerically study a degree of the mutual entrainment 
in the glass phase (spurious memory retrieval).
It is numerically shown in this paper that almost all oscillators synchronize under memory retrieval,
but desynchronize under spurious memory retrieval when setting 
optimal parameters.
Thus, it is possible to determine
whether the recalling process is successful or not using 
information about the synchrony/asynchrony.

In general, when the coupling is sufficiently weak,
the high-dimensional dynamics of a coupled oscillator system
can be reduced to the phase equation \cite{kuramoto,ermentrout}.
Let us consider the following simplified model, 
\begin{eqnarray}
\frac{d \phi_i}{dt} = \omega_i 
 + \sum_{j=1}^{N} J_{ij}\sin(\phi_j-\phi_i+\beta_{ij}), 
 \label{eq.model}
\end{eqnarray}
where $N$ is the total number of oscillators, $\phi_i$ is the 
phase of the $i-$th oscillator, and $\omega_i$ is the natural frequency 
assumed to be randomly distributed over the whole population 
with a density denoted by the symmetric distribution $g(\omega)$,
i.e., $g(\omega) = g(-\omega)$. 
Note that the average of $\omega_i$ may be set to zero without
loss of generality.
The theory presented below can be easily extended to treat the system with 
other any distribution $g(\omega)$.
$J_{ij}$ and $\beta_{ij}$ denote an amplitude of a synaptic weight 
and a synaptic delay, respectively.
In order to investigate the nature of frustrated non-equilibrium systems,
we have selected the following generalized Hebb learning rule \cite{okuda}
to determine $J_{ij}$ and $\beta_{ij}$,
\begin{equation}
 C_{ij} 
  = J_{ij} \exp(i \beta_{ij}) 
  = \frac{1}{N} \sum_{\mu=1}^{p}
                 \xi_{i}^{\mu} {\overline\xi}_{j}^{\mu}, \quad
		 \xi_i^{\mu} = \exp( i \theta_i^{\mu}),
\end{equation}
as their typical example. $\{ \theta_i^\mu \}_{i=1, \cdots, N, \mu=1, \cdots, p}$
are the phase patterns to be stored in the network and are
assigned to random numbers with a uniform probability in $[0, 2\pi]$.
Here, we define a parameter $\alpha$ (loading rate) such that $\alpha=p/N$.
In the equilibrium limit of this model, that is, $g(\omega)=\delta(\omega)$, 
the storage capacity given by $\alpha_c = 0.038$ \cite{cook}.

We put $s_i = \exp(i \phi_i)$ for the sake of simplicity.
The order parameter $m^\mu$, which measures the overlap between 
system state $s_i$ and embedded pattern $\mbox{\boldmath
$\xi$}^\mu$, is defined as 
\begin{eqnarray}
m^\mu = \frac{1}{N} \sum_{j=1}^{N} {\overline\xi}_j^\mu s_j.
\end{eqnarray}
We obtain order parameter equations of the present system
by applying the following manipulations.
\begin{itemize}
 \item First, assuming a self-consistent local field
       for each of oscillators,
       a distribution of $s_i$ under $g(\omega_i)$ 
       is formally derived by the Kuramoto theory.
 \item Second, we estimate the contribution of randomness,
       that is, the uncondensed patterns in the present case
       by the SCSNA,
       and determine the local field in a self-consistent manner.
 \item Finally, the order parameter equations
       are obtained using the self-consistent local field.
\end{itemize}
A detailed derivation of the present theory
will be discussed elsewhere.
Here, we assume $m^1 = O(1)$ and $m^\mu = O(1/\sqrt{N})$ 
for $\mu>1$ (uncondensed patters).
Then, the following two dimensional equations for the order parameters
are obtained,
\begin{eqnarray}
m &=& \left<\left< X(x_1,x_2,\xi) \right>\right>_{x_1, x_2,\xi}, \label{eq2:ope1}\\
U &=& \left<\left< F_1(x_1,x_2,\xi) \right>\right>_{x_1, x_2, \xi}, \label{eq2:ope6}  
\end{eqnarray}
where $\left<\left< \cdots \right>\right>_{x_1, x_2,\xi}$ is taken 
to mean the Gaussian average over $x_1, x_2$ and condensed pattern 
$\xi^1$, $\left<\left< \cdots \right>\right>_{x_1, x_2,\xi} = \left<\int
\int Dx_1 Dx_2\cdots \right>_{\xi}$.
The pattern superscripts $1$ of $m$ are omitted for
brevity. 
The self-consistent mean field $\tilde{h}$, $X$, $F_1$
and the Gaussian measure $Dx_1 Dx_2$ 
are expressed as follows,
\begin{eqnarray}
& &Dx_1 Dx_2=\frac{dx_1 dx_2}{2 \pi \rho^2} 
\exp\left(-\frac{x_1^2+x_2^2}{2 \rho^2} \right),\\
& &\rho^2 = \frac{\alpha }{2 (1-U)^2}, \hspace{0.3cm} \tilde{h}
= \xi m + x_1+i x_2, \\
& &X(x_1,x_2,\xi) = \tilde{h} \int_{-1}^{1} dx  g\left(|\tilde{h}| x \right) \sqrt{1-x^2},\\
& &F_1(x_1,x_2,\xi)=\nonumber \\
& &\hspace{0.5cm}  \int_{-1}^{1}dx 
\left( g \left(|\tilde{h}| x \right)+ \frac{|\tilde{h}|}{2} x g' \left(|\tilde{h}| x \right) \right)
\sqrt{1-x^2}.
\end{eqnarray}
Here, $U$ corresponds to the susceptibility,
which measures the sensitivity to external fields. 
A distribution of resultant frequencies $\overline{\omega}$ 
in the memory retrieval state, which is denoted as $p(\overline{\omega})$, becomes
\begin{eqnarray}
p(\overline{\omega}) &=& r \delta(\overline{\omega})+
\int Dx_1 Dx_2\frac{g\left( \overline{\omega} \sqrt{1 + 
\frac{|h|^2}{\overline{\omega}^2}}\right)}{
\sqrt{1 + \frac{|h|^2}{\overline{\omega}^2}}}, \label{eq:dis} \\
r &=&\int Dx_1 Dx_2 |\tilde{h}| \int_{-1}^{1} dx  g\left(|\tilde{h}| x \right),
\end{eqnarray}
where $r$ measures the ratio between the number of synchronous oscillators
and the system size $N$.
We now consider the relationships between the present theory
and the previously proposed theories.
For the equilibrium limit, $g(x)=\delta(x)$, 
we obtain
\begin{eqnarray}
X = \frac{\tilde{h}}{|\tilde{h}|},\hspace{0.2cm}
F_1 = \frac{1}{2|\tilde{h}|},\hspace{0.2cm} 
p(\overline{\omega}) = \delta(\overline{\omega}),
\end{eqnarray}
which coincide with the replica theory \cite{cook} and the SCSNA \cite{okuda}.
On the other hand, regarding the uniform-system limit, $\alpha=0$, 
our theory reproduces the Kuramoto theory as 
\begin{eqnarray}
m &=& m \int_{-1}^{1} dx  g\left(|m| x \right) \sqrt{1-x^2}, \\
p(\overline{\omega}) &=& r \delta(\overline{\omega})+
\frac{g\left( \overline{\omega} \sqrt{1 +
\frac{|m|^2}{\overline{\omega}^2}}\right)}{\sqrt{1 +
\frac{|m|^2}{\overline{\omega}^2}}}.\\
r &=& |m| \int_{-1}^{1} dx  g\left(|m| x \right).
\end{eqnarray}
As mentioned before, our theory has made a bridge between 
the equilibrium-frustrated system and the non-equilibrium-uniform system.
In addition, we have presented a systematic way of analytical treatments
for the non-equilibrium random systems.
If $C_{ij}$ is
assigned to random numbers with Gaussian in complex plane, 
${\rm Re} [C_{ij}] \sim {\cal N}(1/N, \alpha/2N)$,
${\rm Im} [C_{ij}] \sim {\cal N}(0, \alpha/2N)$,
the order parameter equation consists with Eq. \ref{eq2:ope1}
under constraint $U=0$.
The detail results will be presented elsewhere.
We can also treat Daido's glass oscillator system 
(with real number interaction) \cite{daido} using the same procedure.

In the following analyses, we choose 
a system with $g(\omega)=(2\pi\sigma^2)^{-1/2} \exp(-\omega^2/2\sigma^2)$.
Fig. \ref{phasediagram}(a) shows 
a phase diagram in the  $(|m|, \sigma, \alpha)$ space, 
which was obtained by numerically solving these order parameter equations.
A cross-section of this curved surface at $\sigma=0$
coincides with the results of the SCSNA\cite{okuda} 
and the Replica theory\cite{cook}.
Furthermore, a cross-section of this curved surface at $\alpha=0$
is equal to a result of the Kuramoto theory\cite{kuramoto}.
Thus, our theory bridges the gap between these theories.
Fig. \ref{phasediagram}(b) shows 
the values of critical memory capacity $\alpha_c$ 
for various values of $\sigma$.  
$\alpha_c$ decreases monotonically as  $\sigma$ increases.
The critical value of $\sigma$ at $\alpha_c=0$ is given 
by $\sigma_c = 0.62$, which coincides with that of the 
SK theory.
Figures \ref{sim}(a),(b),(c), and (d) display 
$|m|$ values vs $\sigma$ for various values of $\alpha$, where the solid curves are obtained
theoretically, and the plots show results obtained by 
numerical simulation.
According to these figures, the theory is in good agreement 
with the simulation results.

Next, we examined the distributions of 
the resultant frequencies $\overline{\omega}_i$
over the whole population of oscillators in the memory retrieval state
and the spurious memory state.
Here, the resultant frequencies $\overline{\omega}_i$ were 
calculated by using the long time average of $d \phi_i/dt$.
The plots in Figures \ref{phase1}(a) and (b) show the values of 
resultant frequencies $\{\overline{\omega}_i\}_{i=1, \cdots, N}$
versus $\sigma$, that is, a bifurcation diagram.
Fig. \ref{phase1}(a) denotes $\overline{\omega}_i$ distributions
in memory retrieval states, and Fig. \ref{phase1}(b) represents 
those in spurious memory states.
These results show that there exists a region $\sigma$ 
that satisfies the two conditions: 
all oscillators mutually synchronize in memory retrieval, 
and oscillators desynchronize in spurious memory retrieval.
Thus, it is possible to determine 
whether the recalling process is successful or not only 
using the information 
about the synchrony/asynchrony, when proper $\sigma$ is given.
In Figures \ref{phase1}(a) and (b), 
all phase values $\phi_i$ continued drifting toward
negative directions due to an offset of scattered $\omega_i$.
In other samples, all $\phi_i$ continued drifting toward positive directions.

We investigated the effect of the system size $N$ 
on the $\overline{\omega}_i$ distribution.
The solid curves in Fig. \ref{cook1} indicate 
the distributions of the resultant frequencies
$p(\overline{\omega})$ in Eq. \ref{eq:dis}.
The histograms in Fig. \ref{cook1} 
show the results from the numerical simulation.
As shown in Figures \ref{cook1}(a)(a')(a''), 
the theory agrees well
with the simulation results in memory retrieval states.
As shown in Figures \ref{cook1}(b)(b')(b''), 
there exists a delta-peak that indicates 
the mutual entrainment of a large population of oscillators
in memory retrieval states.
However, in spurious memory states, the distribution of the average frequency
is gentle compared to the distribution in memory retrieval states, 
which indicates the quasi-entrainment of oscillators\cite{daido}.
According to these figures, we believe the phenomena
mentioned before are invariant to the system size $N$.
The reason why all oscillators mutually synchronize in memory retrieval states,
but desynchronize in spurious states, is because
effective interactions among oscillators 
in memory states (as in ferromagnetic states \cite{kuramoto}) 
are different from interactions in spurious memory states
(as in spin glass states \cite{daido}),
in which the system is strongly frustrated.

The phase description proposed here can be considered as
a minimum model of neural networks 
based on oscillatory activities that is mathematically solvable.
Chaos neural networks yield rich phenomena as discussed here, but 
can not be easily analyzed, except with simulations.
Since the present analysis corresponds to
the replica symmetric approximation,
we have noted that it should be extended 
to the replica symmetry breaking
in order to properly treat the spurious states (spin glass states),
and this remains for a future work.

In the field of neuroscience,
a growing number of researchers
have been interested in
the synchrony of oscillatory neural activities 
because physiological evidence of 
their existence has been obtained in the visual cortex of a cat
\cite{Eckhorn1988,Gray1989}.
Much experimental and theoretical research exists
regarding the functional role of synchronization.
One of the more interesting hypotheses is called 
{\it synchronized population coding}, which was proposed by
Phillips and Singer.
However, its validity is highly controversial \cite{Phillips}.
In this paper, we numerically showed the possibility of
determining if the recalling process is successful or not using
information about the synchrony/asynchrony.
If we consider information processing in brain systems,
the solvable {\it toy} model presented in this paper may be
a good candidate for showing the validity of 
a synchronized population coding in the brain,
and we believe the present analysis may strongly influence 
debate on the functional role of synchrony.

The authors thank Ms. M. Yamana, Mr. M. Yoshioka and Dr. M. Shiino
for their valuable discussions.
This work was partially supported by 
Grants-in-Aid for Scientific Research in 
Priority Area (2) No. 07252219,
Grants-in-Aid for the Encouragement of Young Scientists No. 0782 
and JSPS Research Fellowships for Young Scientists.

\begin{figure}
\caption{(a) Phase diagram in $(|m|, \sigma, \alpha)$ space.
(b) $\sigma$  vs critical memory capacity $\alpha_c$.
}
\label{phasediagram}
\end{figure}

\begin{figure}
\caption{Values of $|m|$ vs $\sigma$ (solid curves theoretically obtained, and plots obtained by numerical simulation). 
(a) $\alpha = 0.0$ ($p=1$, $N=1000$). (b) $\alpha = 0.01$ ($N=2000$). 
(c) $\alpha = 0.02$ ($N=2000$). (d) $\alpha = 0.03$ ($N=2000$).
}
\label{sim}
\end{figure}

\begin{figure}
\caption{Values of resultant frequencies $\{\overline{\omega}_i\}_{i=1, \cdots, N}$
vs $\sigma$, $N=2000$ and $\alpha = 0.0315$.
(a) Memory pattern retrieval. 
(b) Spurious memory pattern retrieval.}
\label{phase1}
\end{figure}

\begin{figure}
\caption{Distribution of resultant frequencies
state. $\alpha = 0.01$, $\sigma = 0.32$.
Solid curves show theoretical results
of $p(\overline{\omega})$ in Eq. \protect{\ref{eq:dis}}, but delta-peaks
at $\overline{\omega}=0$ are not indicated.
Figs. (b)(b')(b'') display detailed distributions of Figs.
(a)(a')(a'') at $\overline{\omega}=0$, respectively.
(a)(b) $N=2000$. $20$ trials. (a')(b') $N=4000$. $12$ trials.
(a'')(b'') $N=8000$. $12$ trials.}
\label{cook1}
\end{figure}


\begin{references}
\bibitem{hopfield} J. J. Hopfield, Proc. Natl. Acad. Sci. USA,
{\bf 79}, 2554 (1982).
\bibitem{sherrigton} D. Sherrington, and S. Kirkpatrick,
Phys. Rev. Let. {\bf 35}[26], 1792 (1975).
\bibitem{cook} J.  Cook, J. Phys. A: Math. Gen., {\bf 22}, 2057 (1989).
\bibitem{kuramoto} Y. Kuramoto, {\it Chemical oscillations, waves
and turbulence} (Springer-Verlag, 1984). 
\bibitem{daido} H. Daido, Phys. Rev. Let., {\bf 68}, 1073 (1992).
\bibitem{daido2} H. Daido, Prog. Theor. Phys. {\bf 77}, 622 (1987).
\bibitem{science}J. J. Hopfield and A. V. M. Hertz, 
Proc. Natl. Acad. Sci. USA, {\bf 92}, 6655 (1995).
\bibitem{ermentrout} G. B. Ermentrout, Journal of Mathematical Biology, 
{\bf 6}, 327 (1981). 
\bibitem{okuda}  K. Okuda, (unpublished).
\bibitem{fukai3}  M. Shiino and T. Fukai, J. Phys. A: Math. Gen.,
{\bf 25}, L375 (1992).
\bibitem{okada} M. Okada, T. Fukai and M. Shiino,
Phy. Rev. E, {\bf 57}, 2095-2103 (1998).
\bibitem{Bray1986}
A. J., Bray, H. Sompolinsky and  C. Yu,
J. Phys. C: Solid State Physics,
{\bf 19}, 6389-6406. C. (1986).
\bibitem{Kuhn1991}
K\"{u}hn, S., B\"{o}s, S., \& van Hemmen, J. L. 
Phys. Rev. A, {\bf 43}, 2084 (1991).
\bibitem{Eckhorn1988} R. Eckhorn, R. Bauer, W. Jordan, M. Brosch,
W. Kruse, M. Munk, and H. J. Reitboeck, Biol. Cybern. {\bf 60},
121 (1988).
\bibitem{Gray1989} C. M. Gray, P. K\"{o}nig, A. K. Engel, and W. Singer, 
Nature, {\bf 338}, 334 (1989).
\bibitem{Phillips} W. A. Phillips and W. Singer, Behavioral and
Brain Science, {\bf 20}, 657 (1997).
\end{references}
\end{document}